\documentclass[fleqn,usenatbib]{mnras}

\usepackage{newtxtext,newtxmath}

\usepackage[T1]{fontenc}
\usepackage{multirow}

\DeclareRobustCommand{\VAN}[3]{#2}
\let\VANthebibliography\thebibliography
\def\thebibliography{\DeclareRobustCommand{\VAN}[3]{##3}\VANthebibliography}

\usepackage{graphicx}	
\usepackage{amsmath}	
\usepackage{supertabular}
\defcitealias{Weltevrede2006}{W06}
\defcitealias{Malofeev2018}{M18}

\title[False detections of FRB]{On the problems of detecting Fast Radio Bursts with the LPA LPI}

\author[S. A. Tyul'bashev et al.]
{E. A. Brylyakova, $^{1}$
{
S. A. Tyul'bashev,$^{1}$ \thanks{E-mail: serg@prao.ru (SAT)}
}
\\
$^{1}$ P.N. Lebedev Physical Institute of the Russian Academy of Sciences, Astro Space Center, Pushchino Radio Astronomy Observatory,\\
Radiotelescopnaya 1a, Moscow reg., Pushchino, 142290, Russia\\ 
}

\date{May 01, 2023}
\pubyear{ , 2023}

\begin{document}
\maketitle

\begin{abstract}
The paper presents the verification of previously published fast radio bursts (FRB) from the work of V.A. Fedorova and A.E. Rodin, detected in the monitoring data of the Large Phased Array (LPA) radio telescope using a search algorithm based on the convolution of data with a scattered pulse pattern. The same 6-channel data (channel width 415 kHz) were used for verification, in which FRBs were detected with dispersion measures of 247, 570 and 1767 pc/cm$^3$. Additional verification of the published FRB was also carried out in 32-channel data (channel width 78 kHz). We have not been able to confirm any published FRB on the signal-to-noise ratios stated in the original paper. The main errors are caused by incorrect identification of the baseline and incorrect estimation of the standard deviations of noise.
\end{abstract}





\section{Introduction}

In 2007, a paper was published in which it was said about the detection of a dispersed pulse in archival data obtained from observations on the 64-meter Parkes radio telescope at a frequency of 1.4 GHz. This was the first Fast Radio Burst (FRB) [1]. It was a short pulse less than 5 ms, similar to that of an ordinary pulsar. The observed dispersion measure ($DM$) for this pulse was 375 pc/cm$^3$. Based on the direction in which the burst was determined and the observed $DM$, one could assume its extragalactic origin. The distance to the burst source was estimated as 600 Mpc. The distance estimate and the observed flux density of 30 Jy at a frequency of 1.4 GHz showed that the luminosity of the burst is orders of magnitude higher than the luminosity of ordinary pulsar pulses. This value of the luminosity estimate indicates that the emission mechanism is different from that of ordinary pulsar pulses.

There are 118 sources in the FRB catalog (https://www.frbcat.org/) [2], the last FRB was cataloged in January 2020. Taking into account the work of the Canadian radio telescope CHIME (https://www.chime-frb.ca/home) [3], the number of detected FRBs is currently more than six hundred. According to the FRB and CHIME catalogs, bursts were observed at frequencies from 0.1 to 1.4 GHz, and at $DM$ from 103.5 to 3038 pc/cm$^3$.

There are many different hypotheses attempting to explain the nature of FRBs: the merger of a pair of neutron stars (NS) [4], the merger of a pair of white dwarfs (WD) [5], the merger of an NS and a black hole (BH) [6], the merger of charged BHs [7], the collapse of NS into BH [8], the presence of a planet orbiting a radio pulsar [9], giant pulses of young pulsars [10], giant bursts/flares of magnetars [11, 12], quantum string collisions [13] and others. The observed properties of FRBs and some of the hypotheses about their origin can also be found in reviews [14, 15]. Such a wide variety of hypotheses about the nature of FRBs indicates that the nature of radio bursts is not clear, and the available observations are still not sufficient to select an unambiguous hypothesis about the origin of FRBs.

All currently detected FRBs can be divided into repeating and non-repeating. It is believed that the nature of non-repeating FRBs is associated with some kind of cataclysmic events, while repeating FRBs can be, for example, a manifestation of magnetar activity.

Since the main part of the found FRBs are non-repeating events, observers always collide the question of the reliability of a "one-time" detection. It would seem that choosing events that have a high signal-to-noise ratio
(S/N), you can get rid of false detections. However, in [16], it was shown how a conventional microwave oven can massively generate "new FRBs" with different $DM$s. Trying to avoid such situations,
the authors of the original methods carefully develop the method of processing observations as applied to observations with specific telescopes.

\begin{table*}
\centering
\caption{Checked events}
\begin{tabular}{c|cc|c|c|c|c|c|c|c}
\hline
\label{tab:tab1}
\multirow{2}{*}{Date} & \multicolumn{2}{l|}{Coordinates (J2000)} & \multirow{2}{*}{\begin{tabular}[c]{@{}l@{}}$DM$, \\ pc/cm$^3$\end{tabular}} & \multirow{2}{*}{S/N} & \multirow{2}{*}{\begin{tabular}[c]{@{}l@{}}S$_{peak}$, \\ Jy\end{tabular}} & \multirow{2}{*}{\begin{tabular}[c]{@{}l@{}}$\tau_{s}$, \\ s\end{tabular}} & \multirow{2}{*}{\begin{tabular}[c]{@{}l@{}}$\tau_{s1}$, \\ s\end{tabular}} & \multirow{2}{*}{\begin{tabular}[c]{@{}l@{}}$\Delta t_{6ch}$, \\ s\end{tabular}} & \multirow{2}{*}{\begin{tabular}[c]{@{}l@{}}$\Delta t_{32ch}$, \\ s\end{tabular}} \\ \cline{2-3}
 & \multicolumn{1}{l|}{alpha} & delta &  &  &  &  &  &  &  \\ \hline
06.06.2017 & \multicolumn{1}{l|}{5h 34m} & 41.7 & 247 4 & 8.3 & 0.54 & 0.275 & 0.63 & 0.62 & 0.11 \\
18.10.2015 & \multicolumn{1}{l|}{5$^h$21$^m$} & 33.1 & 570 5 & 6.2 & 1.4 & 0.275 & 2.8 & 1.44 & 0.26 \\
20.09.2016 & \multicolumn{1}{l|}{5$^h$34$^m$} & 41.7 & 1767 4 & 9.1 & 0.22 & 4.33 & 33.3 & 4.48 & 0.82 \\ \hline
\end{tabular}%
\end{table*}

According to the FRB catalog, the median $DM$ value of detected bursts falls within the range of 500–600 pc/cm$^3$. At frequencies above one gigahertz, where the first bursts were detected, the scattering ($\tau$) of the pulse in the interstellar medium slightly broadens the pulse compared to low frequencies. The pulse broadening due to scattering leads to a decrease in the observed peak flux density (a decrease in the S/N). Thus, at frequencies of 100–150 MHz, the scattering is so strong that there are practically no detections of pulsars with $DM$ > 200 pc/cm$^3$. Pulsars are objects with periodic emission and, in contrast to FRBs, it is possible to accumulate a signal for them, increasing the observed S/N. The absence of detections of pulsars with large $DM$ in the meter wavelength range, despite signal accumulation, indicates lower chances of finding single pulses at large $DM$ in low-frequency observations compared to high-frequency observations. Despite the low chances, attempts to detect FRBs in the meter range have been made. For example, in [17], a search was carried out at a frequency of 145 MHz with the LOFAR radio telescope. In [18], the search was carried out at a frequency of 182 MHz with the MWA telescope. The result of the work was upper estimates for the expected number of FRBs per sky, but no real FRB has ever been found.

In 2019, a paper by Fedorova and Rodin [19] was published, according to which the Large Phased Array (LPA) radio telescope of the P.N. Lebedev Physical Institure of the Russian Academy of Sciences (LPI RAS), which is the main instrument of the Pushchino Radio Astronomy Observatory (PRAO), at a frequency of 111 MHz, 3 bursts were found with $DM$=247, 570, 1767 pc/cm$^3$. At the PRAO seminar, the work caused a large number of criticisms regarding the processing of observations and the reliability of the results obtained. In the present work, we verified the discovered sources using the same data that were used by the authors and repeating the proposed search technique.

\section{CHECKING FOR FOUND FRBs}

The search for pulsed radiation was carried out in the data obtained on the LPA LPI radio telescope. LPA LPI is a meridian–type radio telescope with a filled aperture, which is a flat equidistant array consists of 16,384 wave dipoles. The radiation pattern size is approximately 0.5$^{\circ} $x1$^{\circ}$, the time of passage of the source through the meridian is 3.5 min/cos$\delta$ ($\delta$ is the declination) at half power. The central observation frequency is 110.3 MHz, the reception bandwidth is 2.5 MHz. Daily and round-the-clock monitoring in 96 beams of the telescope began in August 2014, but partial monitoring (not for all beams and not for all dates) has been carried out since 2012.

Data is synchronously recorded in two modes: six channels with a channel width of 415 kHz and a sampling time of one point $\Delta$t = 100 ms; 32 channels with a channel width of 78 kHz and a sampling time of $\Delta$t = 12.5 ms. To search for FRBs, Fedorova and Rodin [19] used data with a low time-frequency resolution.

The search for new FRBs in [19] was carried out
according to the following scheme:

– the baseline (the background of the Galaxy) was subtracted from the original hourly data file, representing the initial data, which, according to [19], were smoothed by the median filter. The median filter is not described in the work, and we used the usual definition. An array is taken from a data file. The points are arranged in ascending or descending order. The output value after running the median filter is the point in the middle position. A shift is made in the original array by a length equal to the length of the array in which the median value was searched, and the procedure is repeated. As a result, we have a set of points that can be connected by segments. It is assumed that each the segment describes the base line along the length of the segment;

– for the search, a twenty-minute interval was taken, the coordinate of the center of which in right ascension was 5$^h$ 32$^m$ (the coordinate of the repeating FRB 121102);

– $DM$ was enumerated with a step of 50 pc/cm$^3$ in the range of 0 $\leq$ $DM \leq$ 3000 pc/cm$^3$, and after adding the frequency channels with the expected $DM$, convolution was made with the template. The template is a model pulse broadened in the receiver band according to the test $DM$ = 360 pc/cm$^3$ and convolved with an exponential function (thin screen assumption) with a scattering scale $\tau_{s}$ = 1s, assuming $\tau_{s} \sim DM^{-2.2}$;

– upon detection of an FRB candidate, convolution was performed with pulses corresponding to different $DM$s, and it was considered that the $DM$ was determined correctly when the resulting S/N ratio maximum. Here and below, we defined the S/N as the ratio of the value of the maximum to be checked in the array of points on the expected $DM$ of the FRB candidate to the root-mean-square deviations of the noise in this array on the $DM$ = 0 pc/cm$^3$. The transfer function for taking into account scattering for the template was calculated using the formula:

\begin{equation}
    F(t) = \frac{1}{t_{s}}\int_{t}^{0}exp(-\frac{t}{t_{s}})\delta(t - \tau)d\tau,
	\label{eq:eq_1}
\end{equation}
where $t_{s}$ is the pulse broadening due to scattering according to the model [20] for the tested $DM$.

\begin{figure*}
\begin{center}
	\includegraphics[width=0.8\textwidth]{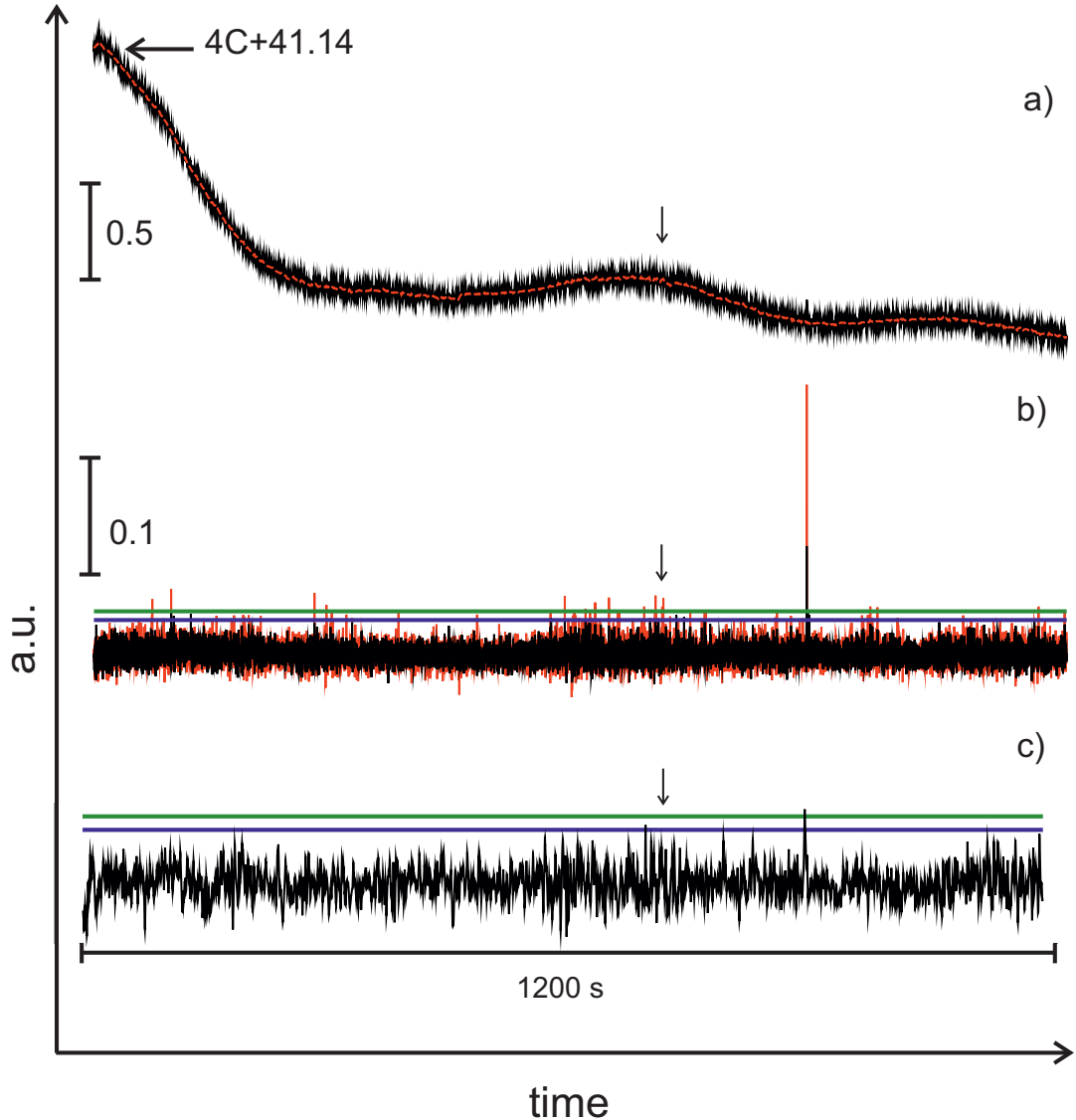}
    \caption{On fig.(a) 20-minute raw data record on June 6, 2017 without $DM$ compensation (black color). The background signal being subtracted is shown in red. At the beginning of the record, the peak of the 4C+41.14 source is visible with a flux density of 16.1 Jy at a frequency of 102.5 MHz according to the catalog http://astro.prao.ru/db/; (b) data after baseline subtraction before (red) and after (black) compensation with $DM$ = 247 pc/cm$^3$; (c) data after passing the convolution procedure with the scattering function according to Kuzmin et al. [20] in 6-channel data}
    \label{fig:fig1}
\end{center}
\end{figure*}

The data taken in the interval from July 2012 to May 2018 were processed for two directions in the sky. For each direction, 355 hours of recording were accumulated over 6 years. We partially reproduce Table. 1 from article [19] with data on the found FRBs. Two columns with expected FRB redshifts and pulse energies have been removed from the original table. The remaining columns of the table sequentially present: the dates of FRB detection, coordinates, their $DM$ and S/N estimates, the estimate of the peak flux density ($S_{peak}$), and the estimate of the characteristic scattering time ($\tau_{s}$) according to the observations of Fedorova and Rodin [19]. The adjacent column gives our estimate of the expected scattering ($\tau_{s1}$) for putative $DM$ FRB candidates calculated using the empirical model of Kuzmin et al. [20]. In this model, pulsars with $DM$ < 200 pc/cm$^3$ were considered, and it was found that Fedorova and Rodin [19] assumed that on $\tau_{s}(DM) = 60\times(DM/100)^{2.2\pm 0.1}$.
At higher values of $DM$, the slope of the scattering value versus $DM$ does not change. Note that for pulsars the dependence can be even steeper (see [21, 22]), but for extragalactic sources the formal application of scattering from [20] can be justified [23]. In columns 9 and 10 we present the intra-channel broadening of the pulse, i.e. its dispersion smearing, in frequency channels for 6 ($\Delta t_{6ch}$) and 32 ($\Delta t_{32ch}$) channel observations, at the $DM$ value indicated in column 4 and determined by the formula:

\begin{equation}
    \Delta t = 4.15\times10^6(\frac{1}{\nu_1^2} - \frac{1}{\nu_2^2})DM,
	\label{eq:eq_2}
\end{equation}
where $\Delta t$ is the time delay of the signal due to passage through the medium in milliseconds, $\nu_1$ and $\nu_2$ are the midpoints of the frequency channels, expressed in MHz.

Below we consider the processing of the event of 06.06.2017, in which an FRB candidate was found that has the smallest dispersion measure ($DM$ = 247 pc/cm$^3$), and, consequently, the minimum line of sight scattering and intrachannel pulse broadening. The coordinates of the FRBs found by the authors of the original paper, Fedorova and Rodin [19], are given in right ascension with an accuracy of one time minute, so the expected location can be $\pm$1 min from the event coordinate. Visual search was carried out by us on the entire data segment, i.e. $\pm$10 min from the event coordinate. After compensating the gain in the frequency channels according to the calibration signal (see details in [24]), the recording was shifted in individual channels, taking into account $DM$, then the channels were combined and the subsequent cross-correlation with the template $\tau_{s1}$, the value of which is given in Table~\ref{tab:tab1}.

On fig. 1a–1c show the stages of processing. On fig. 1a shows raw data, the center corresponds to the coordinate 5$^h$32$^m$. According to [19], the baseline is the original data smoothed by the median. The median step is not given in the original work. When choosing the median step, we were guided by the considerations that if the median step is chosen less than the pulse scattering time, then the pulse in the processed record can be strongly smoothed or destroy altogether. If the median step is chosen many times larger than the scattering time, then after subtracting the baseline, poorly subtracted discrete sources and details of the background of the Galaxy may remain in the remaining record. So, according to Table~\ref{tab:tab1}, the expected scattering of an FRB having $DM$ = 1767 pc/cm$^3$ is almost half a minute, and the LPA half power beam size is $\sim$ 3.5 min. Selecting a median step of 3 min (six times larger $\tau_{s1}$) will result in to the fact that discrete sources will be visible in the recording. It is obvious that the median step must be chosen depending on the expected amount of scattering. When searching for FRB, the scattering is \textit{a priori} unknown, the magnitude of which depends on $DM$. Therefore, it is necessary to enumerate all possible values of the median step corresponding to different scatterings. In the present work, we test candidates who have their $DM$ known and an estimate of the scattering magnitude. For the median step, we chose twice the size of the estimate $\tau_{s1}$ according to the model of Kuzmin et al. [20]. For the tested FRB with $DM$ = 247 pc/cm$^3$, the median step was chosen to be 1.2s (12 points). In all figures, the horizontal axis shows time, and the vertical axis shows the flux density in conventional units. The blue and green lines in the figures show the levels 4$\sigma_{n}$ and 5$\sigma_{n}$. Vertical segments show height values in conventional units. The arrow indicates the expected location of the source from Table~\ref{tab:tab1}.

On Fig.1b presents the data of Fig.1a after baseline subtraction before (red) and after (black) $DM$ compensation. Separate peak signals are visible, the maxima of which are higher before, and not after, dispersion compensation. That is, these peak signals are ordinary impulse noise having $DM$ = 0 pc/cm$^3$. On Fig.1c shows the data after convolution. It can be seen that some of the impulse signals after dispersion compensation and addition of frequency channels have levels higher than 4$\sigma_{n}$, but less than 5$\sigma_{n}$ (blue (lower) and green (upper) lines in the figure). However, verification shows that all these signals are associated with impulse noise, the height of which at $DM$ = 0 pc/cm$^3$ higher than their height after the addition of the channels, taking into account $DM$ = 247 pc/cm$^3$. The remaining peak signals visible after convolution with the scattered signal are less than 4$\sigma_{n}$.

We also ran parallel processing on 32-channel data. The final patterns after convolution for 6- and 32-channel data do not match. If there are real sources in the records, the sensitivity in data with high time-frequency resolution should almost always be higher than in data with low time-frequency resolution. In data with a channel width of 415 kHz (6-channel data), the broadening within the frequency channel due to the dispersion lag for the case under study is about 5.4 times greater than in 32-channel data (channel width of 78 kHz), and, therefore, the S/N in 32-channel data can grow to $\sqrt{5.4} \simeq 2.5$ once. In other words, if the original signal was narrow and its scattering was less dispersion smearing in a frequency channel, the narrower the channel, the greater the gain in sensitivity for 32-channel data, all other things being equal. If the scattered pulse width was greater than the dispersion smearing in the channel, then the gain in sensitivity will not, but with the same averaging of data over time, the pulse profiles should repeat each other. However, a comparison of the processing of 6- and 32-channel data shows that no increase in S/N, and in general any pronounced signal at S/N levels > 4, is visible in the records. The signals for 6- and 32-channel data do not repeat each other, and we can say that in the considered case we are talking about a purely noise process.

When checking the event of October 18, 2015, where the authors detected a signal with $DM$ = 570 pc/cm$^3$, a problem arose related to the right ascension coordinates of the event used by Fedorova and Rodin [19]. In this work, 20 min of the record with the center at 5$^h$32$^m$ were studied, and therefore the extreme points of the record should have coordinates 5$^h$22$^m$ and 5$^h$42$^m$. In this case, the coordinate of the found event is 5$^h$21$^m$. It is possible that this is a typo, it is possible that the area studied is slightly larger than the 20 min. specified in the article. We checked both the area 5$^h$22$^m$ - 5$^h$42$^m$ and the area around 5$^h$21$^m$. Figures 2a–2f show the event for October 18, 2015, where we managed to detect a signal that was as close as possible to the signal received from the authors. It is located at coordinates close to 5$^h$21$^m$. The baseline was subtracted by straight line segments, the length of which (5.6s, 56 points) was determined by the double value of the expected scattering time (see Table~\ref{tab:tab1}).

Figures 2d–2f are similar in terms of the location of the maxima and the general shape. The left side shows 6-channel data, the middle one shows 32-channel data, and the right side shows a fragment of the original figure with the supposed FRB detection in [19]. However, from Fig.2c (see arrow) also shows that the S/N of these peaks is much smaller than 4$\sigma_{n}$, which indicates the absence of significant signals.

For the event on September 20, 2016 (Fig.3) with $DM$ = 1767 pc/cm$^3$, the pulse in the channel should spread up to 33s due to scattering. Subtracting the data smoothed by the median with a step of 1 min leads to the appearance of trends in the record (Fig.3a). On Fig.3b the record after convolution with the scattered signal is shown. We failed to find a signal similar to the signal given in [19].

Thus, our estimates give S/N < 4 for all events from [19]. We believe that this indicates the absence of real detections.

\begin{figure}
\begin{center}
	\includegraphics[width=\columnwidth]{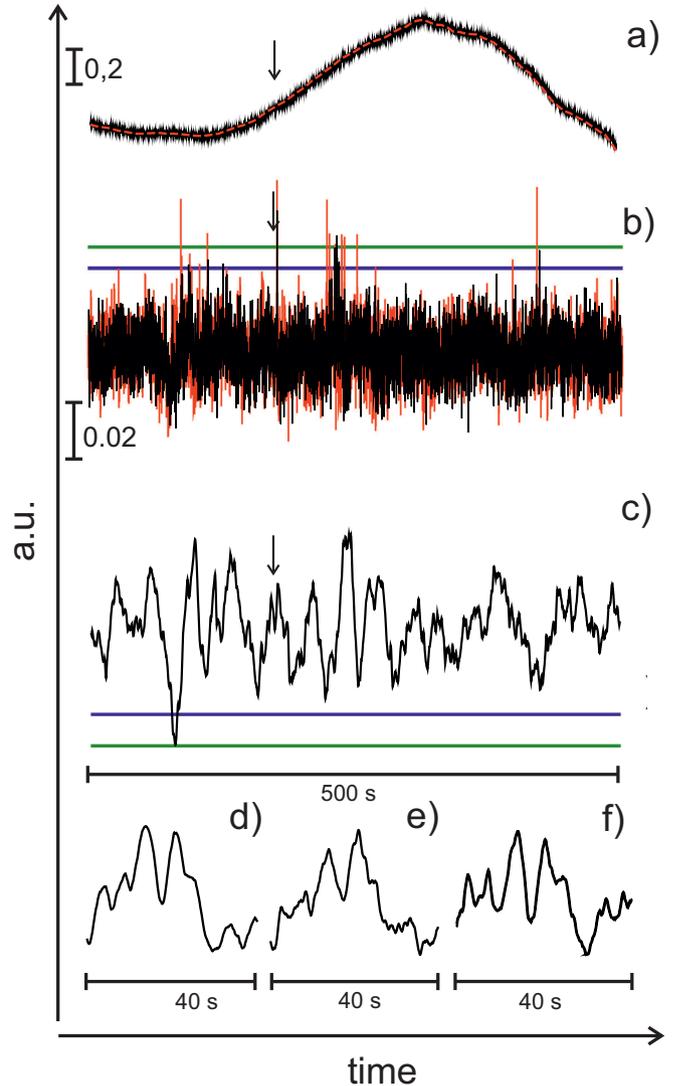}
    \caption{On fig. (a) black color shows approximately 8 min of raw data on 10/18/2015 with the center. The baseline being subtracted is shown in red; (b) signal after baseline subtraction and compensation $DM$ = 570 pc/cm$^3$;
    (c) signal after convolution with scattered pulse and summation of all frequency channels. The arrow points to the detected signal, which coincides with the signal from [19]; (d) and (e) signals after processing of 6- and 32-channel
    data most similar to the signal from [19], a fragment of which is shown in Fig. (e).}
    \label{fig:fig2}
\end{center}
\end{figure}

\begin{figure}
\begin{center}
    \includegraphics[width=0.8\columnwidth]{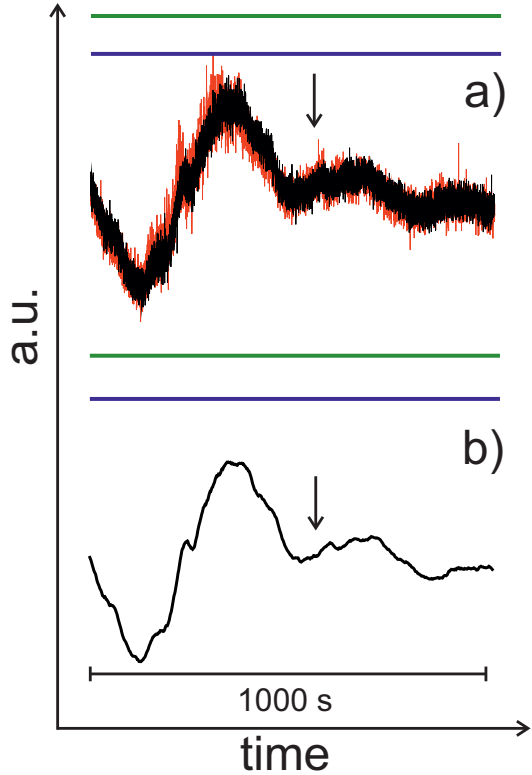}
    \caption{On fig. (a) shows the results of baseline subtraction before (red) and after (black) accounting for smearing in the band with $DM$ = 1767 pc/cm$^3$ of the FRB candidate on September 20, 2016; (b) signal after convolution with a scattered pulse.}
    \label{fig:fig3}
\end{center}
\end{figure}

\section{DISCUSSION of THE RESULTS} 

In the previous paragraph, we considered the main, in our opinion, sign of the discovery of a new FRB - this is its S/N. The S/N level at which the probability of a false detection becomes negligible depends on the number of cases considered. In radio astronomy, the criterion S/N = 6–7 is adopted, when the probability of false detection becomes
negligible low. An important condition in determining the S/N is the correct subtraction of the average level (baseline). It was shown that at the coordinates corresponding to the detected FRBs, there are no signals with a S/N level > 4, which contradicts the declared S/N values from 6.1 to 9.2 [19]. The signal levels found by us clearly indicate the low reliability of the detection of new FRBs.

We do not consider in our paper all the comments made at the PRAO seminars, but we note some of them.

– The dispersion delay lines in the figures are smooth, and there is a feeling of a large number of used frequency channels. What is it connected with?

– On the given convolutions of the original signal with a scattered profile, the baseline (average signal level) is not indicated, which does not allow visually assessing the significance of the event. For example, for a candidate with $DM$ = 247 pc/cm$^3$ (see [19], Fig. 8), there is a minimum (count 50) of the same scale in front of the maximum (count 80). There is a feeling that the conducted noise convolution with scattered model signal simply redistributed the energy
noise track and gave a minimum and maximum with the same energy.

– When searching for dispersed pulses for each direction under study, many iterations are made on different $DM$s, and random noise arrays are obtained with each enumeration. These enumerations were made for observations made daily over an interval of 6 years. How many iterations during the search were actually made, and what is the probability of a statistically significant detection of a false (noise) signal on different S/Ns?

– Why, for all FRB candidates, the given scattering estimates are much less than those expected from the empirical relationship of the scattering value from $DM$ according to [20], which was taken
authors as a basis for scattering estimates? Why does the duration (half-width) of pulses for the found sources not correspond to empirical models of pulse scattering in the interstellar medium?

– Why are high time-frequency resolution data (32 channels) that are recorded in parallel with low time-frequency resolution data (6 channels) the S/N of found sources less for the same time averaging? The question was raised by the fact that in the 32-channel data, the intrachannel dispersion smoothing is 5 times smaller, which should lead to larger values of the observed burst S/N, at least for a source with $DM$ = 247 pc/cm$^3$.

– It seems strange that two events with $DM$ = 247 and 1767 pc/cm$^3$
  coincided in coordinates.

– According to [19], FRB 121102 ($DM$ = 570 pc/cm$^3$) could be detected in the side lobe of the LPA LPI. The maximum of the first side lobe is approximately 4\% of the energy in the main lobe. Therefore, if this pulse were observed in the main lobe, its signal-to-noise ratio would be S/N > 6.2 $\times$ 25 >  150. As of 2021, more than 1600 pulses have been recorded for FRB 121102 [25]. What is the probability that in the sidelobe, where there is an insignificant part of the total pulse energy, a pulse was registered, and at the same time, not a single pulse was detected in the main lobe for 6 years? The alternative explanation of the authors suggests the possible detection of some other FRB. In this case, it is surprising that its $DM$ matches with great accuracy the $DM$ of the known repetitive FRB.

As it turned out at the seminars, the figures called dynamic spectra in [19] are not them in the generally accepted sense. The convolution of the initial data with the scattering pattern acts as the flux density in the frequency channel. To get a "beautiful" picture the authors used digital filters not described in the paper, which smoothed the image, giving a false impression of a large number of observed frequency channels and continuity of the dispersion delay line. Most likely, these are non-linear filters. Upon receipt a very similar profile in 6-channel data for a pulse with $DM$ = 570 pc/cm$^3$, we have not been able to obtain a similar picture of the dynamic spectrum.

Let's give an answer to one of the questions asked at the seminars, namely, to the question about the probability of a false signal detection. Fortunately, there is enough data in [19] to provide an answer. We assume that the noise is distributed according to the normal law. The number of iterations in the search for a transient for each beam of the LPA LPI can be determined as 355h $\times$ 3600 (seconds per hour) $\times$ 10 (counts per second) $\times$ 60 (sorted $DM$) = 766.8 million independent and verifiable events in which the authors tried to find dispersed impulses. Even at a level above 4$\sigma_{n}$, 48,571 false events will be detected, at a level above 5$\sigma_{n}$ - 439 events, at a level above 6$\sigma_{n}$ - 1 event. We have given estimates of the probability of false detection of an impulsive signal for ideal white noise. Taking into account the noise recorded at the LPA LPI both in the frequency and in the time domain, and having, among other things, explicit dispersion delays of signals [26, 27], the obtained estimates of the number of expected false events should be taken as lower estimates. According to the previous paragraph, the estimates of S/N < 4 for all tested events. Therefore, we believe that all three "found" events are false detections.

One of the unpleasant moments that appeared when trying to repeat the technique [19] on the LPA LPI data was the choice of the median step used to obtain smoothed data and subsequent subtraction of the baseline. In [19], nothing is said about the step of the median for background subtraction, as well as about the method for determining the S/N ratio. We wrote above that we used twice the scattering value on the tested $DM$. However, observations [20–22] show that the scattering can be an order of magnitude larger than in the model [20]. In this case, the step of the median must also change accordingly. If it does not change (not increase), you can lose pulse. Thus, when searching for an FRB with $DM$ = 1767 pc/cm$^3$, the median step can be up to 33s$\times$10 = 5.5 minutes. The median step of 5.5 min is almost equal to the time it takes the source to pass through the antenna pattern. Therefore, with such a chosen median step, discrete sources will not be subtracted and will remain in the record. This means that when searching for FRB candidates in the LPA data, it is necessary to restrict ourselves to such dispersion measures, the median step of which is such that discrete sources disappear after subtracting the baseline.

For the candidate with $DM$ = 247 pc/cm$^3$, we repeated the processing with a median step of 12s (120 points), which is 20 times the expected scattering time. FRB was not found. For the candidate with $DM$ = 570 pc/cm$^3$, no reprocessing with a median step corresponding to the larger expected scattering time was performed, because the resulting picture (see Fig.2) almost completely coincided with the picture from [19]. The candidate with $DM$ = 1767 pc/cm$^3$ was not reprocessed for the reasons given in the previous paragraph.

Critical remarks on the reliability of the detection of new FRBs, without consideration of the processing technique, were made to two papers by Fedorova and Rodin in [28], where at low frequencies, a repeating FRB 20180916B was detected.

In 2021, work [29] was published, where the S/Ns of the cases we considered were changed. In particular, for the case with $DM$ = 570 pc/cm$^3$, instead of S/N = 6.2, the new value is S/N = 5.9. However, as shown above, for this case, S/N < 4. No attempts have been made to correct the search technique. The S/N ratio for almost half of the 60 "new FRBs" is less than 6, and for five candidates it is less than 4$\sigma_{n}$. Breaks are observed in some of the presented dynamic spectra. Of course, there may be real FRBs among the 60 published cases. It is necessary to carry out a complete audit of all candidates. This is beyond the scope of this work, which is only devoted to testing the search methodology and the first three candidates discovered.
  
\section{CONCLUSIONS}

When searching for dispersed pulses, usually the evidence for the discovery of a new source is an estimate of S/N = 6–7 or more. S/N is defined as the amplitude of the peak signal divided by the standard deviation noise. The standard deviation of noise, other things being equal, is the smaller, the better the baseline is subtracted. The baseline is a set of line segments. The length of these segments is determined by the amount of momentum scattering in the interstellar and intergalactic medium. If the dispersion measure is high, then the length of the segments used to subtract the background can be comparable to the duration of the passage of a discrete source through the radiation pattern of the LPA LPI. Thus, the search for FRBs on large $DM$s at the LPA LPI radio telescope is automatically limited. After detecting signals with S/N > 6–7, it is necessary to check the source by indirect signs. It should be detected in one or two LPA beams. The dynamic spectrum should show a pronounced dispersion delay line.

When checking the FRB candidates found in [19], no signals with S/N > 4 were found, which indicates that all these detections are false. It is possible that the authors, in order to estimate the standard deviations, took a short segment of the processed record, on which they see something similar to an pulse, and calculated the standard deviations on a small piece of the record outside the impulse. The processed record with a large number of significant points shows that the standard deviations of the noise are much larger, and the determined S/N value is much smaller. 

Unfortunately, the method of Fedorova and Rodin [19] regarding the choice of the step for the median filter and the choice of points for estimating standard deviations is written so ambiguously that we had to be guided by common sense meaning when trying to repeat the processing technique. We managed to "discover" only one of the three "FRBs" presented in the work. Judging by the noise track in Fig. 2c and the value of S/N = 1, equal to the distance between the lines drawn at S/N = 4 and 5, for this "FRB" S/N < 2.

In this work, the question of the fundamental possibility or impossibility of detecting FRBs at the LPA LPI remained unanswered. Here are some considerations that speak in favor of possible FRB detections. The main factor hindering the detection of FRBs is the scattering of pulse in the interstellar medium. It is known that the scattering for any chosen $DM$, according to observations in the meter wavelength range, can vary by more than an order of magnitude, both up and down from the average value [20]. The average values extracted from Fig.2 of this work show that the scattering $\tau_s$=10, 30, 200 ms on dispersion measures $DM$ = 50, 100, 200 pc/cm$^3$, but can be less than 1, 10, 100 ms on the same dispersion measures. According to the FRB catalog [2], after excluding the "Pushchino" FRBs, the widths of the found pulses are in the range from 0.08 to 34 ms, and the median value is 1.9 ms. Recall that the sampling time in the monitoring survey conducted at the LPA LPI is 12.5 ms. In the same catalog, 11 FRBs (approximately 10\% of the list of all FRB sources in the catalog) have an observed $DM$ < 200 pc/cm$^3$, and the minimum is 103.5 pc/cm$^3$. The CHIME radio telescope found an FRB with $DM$ = 81.82 pc/cm$^3$ [30], which is not yet included into the FRB catalog. Thus, FRBs have already been found, the pulse widths and dispersion measures of which indicate the possibility of their detection in observations at the LPA LPI, if, of course, the sensitivity of the instrument is sufficient.

In the work on the search for rotating transients, carried out in the PRAO [31], Table 1 and Figure 1 shows the sensitivity of the LPA when searching for single pulses in comparison with the sensitivity of other large instruments on which FRBs have already been detected. The threshold sensitivity was recalculated to a frequency of 111 MHz assuming a spectral index ($\alpha = -1.5 (S \sim \nu^{\alpha}$)). According to this table, the threshold sensitivities in the search for pulsed signals with S/N = 7 for telescopes with a mirror diameter of 64, 100, 300 m (Parks, Effelsberg, Arecibo) and LPA antennas are 4, 0.55, 0.5, and
2.1 Jy. The sensitivity of 2.1 Jy in the search for pulsed radiation at the LPA has been confirmed by practical detections of new RRATs and pulses from known pulsars [31]. The weakest detected pulses have a flux density of approximately 2 Jy. Pulses have been detected from 115 known pulsars and 46 new RRATs (https://bsa-analytics.prao.ru/en/transients/rrat/). Thus, the confirmed sensitivity of LPA in the search for FRB is at a level comparable to the sensitivity levels of some of the the best radio telescopes in the world. Finally, the detection of the repeating FRB 20180916B in the frequency range 110–190 MHz with the LOFAR radio telescope [32] once again indicates that detections of FRBs with $DM$ < 200–300 pc/cm$^3$ at the LPA LPI is a matter of time and competent processing of observations.

\section*{Acknowledgements}
The authors express their gratitude to the staff of the observatory V.M. Malofeev, T.V. Smirnova, V.A. Potapov for preliminary reading of the manuscript and a number of comments that made it possible to improve the text, as well as L.B. Potapova for assistance in the design of figures.

The study was carried out at the expense of a grant Russian Science Foundation 22-12-00236, https:// rscf.ru/ project/ 22- 12- 00236/.

\section*{Bibliography}
1. D. R. Lorimer, M. Bailes, M. A. McLaughlin, D. J. Narkevic, and F. Crawford, Science (Washington,
DC, U. S.) {\bf 318}, 777 (2007).

2. E. Petroff, E. D. Barr, A. Jameson, E. F. Keane, et al., Publ. Astron. Soc. Austral. {\bf 33}, e045 (2016).

3. M. Amiri, B. C. Andersen, K. Bandura, S. Berger, et al., Astrophys. J. Suppl. {\bf 257}, 59 (2021).

4. J. Abadie, B. P. Abbott, R. Abbott, M. Abernathy, et al., Class. Quantum Grav. {\bf 27}, 173001 (2010).

5. K. Kashiyama, K. Ioka, and P. Meszaros, Astrophys. J. Suppl. {\bf 776}, L39 (2013).

6. C. M. F. Mingarelli, J. Levin, and T. J. W. Lazio, Astrophys. J. Lett. {\bf 814}, L20 (2015).

7. B. Zhang, Astrophys. J. Lett. {\bf 827}, L31 (2016).

8. H. Falcke and L. Rezzolla, Astron. Astrophys. {\bf 562},
A137 (2014).

9. F. Mottez and P. Zarka, Astron. Astrophys. {\bf 569}, A86
(2014).

10. L. Connor, J. Sievers, and U.-L. Pen, Mon. Not. R.
Astron. Soc. {\bf 458}, L19 (2016).

11. S. B. Popov, in Evolution of Cosmic Objects through
Their Physical Activity, Proceeding of the Conference dedicated
to Viktor Ambartsumian’s 100th Anniversary, September
15–18, 2008, Yerevan and Byurakan, Armenia,
Ed. by H. A. Harutyunian, A. M. Mickaelian, and
Y. Terzian (Gitutyun Publ. House of NAS RA, Yerevan,
2010), p. 105.

12. S. R. Kulkarni, E. O. Ofek, J. D. Neill, Z. Zheng, and
M. Juric, Astrophys. J. {\bf 797}, 70 (2014).

13. L. V. Zadorozhna, Adv. Astron. Space Phys. {\bf 5}, 43
(2015).

14. S. B. Popov, K. A. Postnov, and M. S. Pshirkov, Phys.
Usp. {\bf 61}, 965 (2018).

15. J. M. Cordes and S. Chatterjee, Ann. Rev. Astron. Astrophys.
{\bf 57}, 417 (2019).

16. E. Petroff, E. F. Keane, E. D. Barr, J. E. Reynolds,
et al., Mon. Not. R. Astron. Soc. {\bf 451}, 3933 (2015).

17. A. Karastergiou, J. Chennamangalam, W. Armour,
C. Williams, et al., Mon. Not. R. Astron. Soc. {\bf 452},
1254 (2015).

18. A. Rowlinson, M. E. Bell, T. Murphy, C. M. Trott,
et al., Mon. Not. R. Astron. Soc. {\bf 458}, 3506 (2016).

19. V. A. Fedorova and A. E. Rodin, Astron. Rep. {\bf 63}, 39
(2019).

20. A. D. Kuz’min, B. Y. Losovskii, and K. A. Lapaev, Astron.
Rep. {\bf 51}, 615 (2007).

21. N. D. R. Bhat, J. M. Cordes, F. Camilo, D. J. Nice, and
D. R. Lorimer, Astrophys. J. {\bf 605}, 759 (2004).

22. A. V. Pynzar’ and V. I. Shishov, Astron. Rep. {\bf 52}, 623
(2008).

23. M. Amiri, K. Bandura, M. Bhardwaj, P. Boubel, et al.,
Nature (London, U. K.) {\bf 566}, 230 (2019).

24. S. A. Tyul’bashev, P. Yu. Golysheva, V. S. Tyul’bashev,
and I. A. Subaev, Astron. Rep. {\bf 63}, 920 (2019).

25. D. Li, P. Wang, W. W. Zhu, B. Zhang, et al., Nature
(London, U. K.) {\bf 598}, 267 (2021).

26. V. A. Samodurov, S. A. Tyul’bashev, M. O. Toropov,
and S. V. Logvinenko, Astron. Rep. {\bf 66}, 341 (2022).

27. S. A. Tyul’bashev, D. V. Pervukhin, M. A. Kitaeva,
G. E. Tyul’basheva, E. A. Brylyakova, and A. V. Chernosov,
arXiv: 2204.02025 [astro-ph.HE] (2022).

28. Z. Pleunis, D. Michilli, C. G. Bassa, J. W. T. Hessels,
et al., Astrophys. J. Lett. {\bf 911}, L3 (2021).

29. V. A. Fedorova and A. E. Rodin, Astron. Rep. {\bf 65}, 776
(2021).

30. M. Bhardwaj, B. M. Gaensler, V. M. Kaspi, T. L. Landecker,
et al., Astrophys. J. Lett. {\bf 910}, L18 (2021).

31. S. A. Tyul’bashev, V. S. Tyul’bashev, and V. M. Malofeev,
Astron. Astrophys. {\bf 618}, A70 (2018).

32. I. Pastor-Marazuela, L. Connor, J. van Leeuwen,
Y. Maan, et al., Nature (London, U. K.) {\bf 596}, 505
(2021).

\bibliographystyle{mnras}

\end{document}